\documentclass[twocolumn,aps,prl,amsmath,amssymb,letterpaper,showpacs]{revtex4}

\usepackage{graphicx}
\usepackage{amsmath}
\usepackage{amssymb}
\usepackage{epstopdf}
\usepackage{bm}
\usepackage{dcolumn}

\begin{document}
\title{Nuclear quantum effects in water}
\author{Joseph A. Morrone}
\affiliation{Department of Chemistry, Princeton University \\ Princeton, NJ 08544}

\author{Roberto Car}
\email{rcar@princeton.edu}
\affiliation{Department of Chemistry and Department of Physics, Princeton University \\ Princeton, NJ 08544}

\date{\today}

\begin{abstract}
In this work, a path integral Car-Parrinello molecular dynamics simulation of liquid water is performed. It is found that the inclusion of nuclear quantum effects systematically improves the agreement of first principles simulations of liquid water with experiment.  In addition, the proton momentum distribution is computed utilizing a recently developed open path integral molecular dynamics methodology. It is shown that these results are in good agreement with neutron Compton scattering data for liquid water and ice. 
\end{abstract}
\maketitle

Due to the fundamental importance of water in the physical and biological sciences, understanding its microscopic structure is an issue of longstanding interest.  Elucidating the local environment of the protons is particularly intriguing due to their crucial role in hydrogen bonding.  Nuclear quantum effects have a significant impact on the behavior of water.  This is indicated by the large isotope effects observed in numerous water properties when protons are substituted with deuterium (D) or tritium (T) nuclei. For example, the melting point of heavy water (D$_2$O) is 3.82 K higher than that of regular (H$_2$O) water, and the effect is even more pronounced in tritiated water (T$_2$O)~\cite{greenwood}, providing evidence that quantum effects destabilize the hydrogen bond network.

Recently, the equilibrium state of the protons in water and ice has been probed by neutron Compton scattering experiments~\cite{reiter}. This technique measures the proton momentum distribution~\cite{noreland}, thereby providing complementary information to what is garnered from diffraction techniques that measure the spatial correlations among the nuclear positions~\cite{soper_2000,headgordon2}. Due to the non-commuting character of position and momentum operators in quantum mechanics, the proton momentum distribution is sensitive to the local environment. In particular, the differences in the momentum distribution of the solid and liquid water phases reflect the breaking and distortion of hydrogen bonds that occurs upon melting. In systems such as confined water~\cite{reiternano,reitersilica} and the quantum ferroelectric potassium phosphate~\cite{kdp}, the momentum distribution provides signatures of tunneling and delocalization.

Molecular simulations with quantum nuclei are made feasible by the Feynman path integral representation of the equilibrium density matrix at finite temperature. This approach has been used in conjunction with empirical force fields in studies~\cite{comment2} showing that quantum fluctuations soften the structure of liquid water.  The effect is illustrated by a broadening of the radial distribution functions (RDF) compared to those generated from classical nuclei. Interestingly, these works indicated that quantum nuclei affect the structure in a similar way to a temperature increase in a classical simulation. Recently, empirical force fields have been employed within ``open'' path integral molecular dynamics methodologies to compute the proton momentum distribution in ice and water~\cite{burnham,morrone}.  The calculated distribution, while in agreement with experiment in many respects, did not reproduce the shorter tail that is observed in ice, signaling a lack of transferability of the empirical potentials. The faster decaying ice distribution reflects a red-shift of the OH stretch frequency that is a consequence of the recovery of an intact hydrogen bond network upon freezing. 

To investigate whether this effect can be reproduced in ab-initio simulations, we perform an ``open'' path integral Car-Parrinello molecular dynamics (PI-CPMD)~\cite{cppimd} study of water in the liquid and solid phases. In this approach the nuclear potential energy surface is derived on the fly from the instantaneous ground state of the electrons within Density Functional Theory (DFT). Our study is also motivated by a previous, pioneering PI-CPMD simulation of liquid water~\cite{waterpi03}. This study reached the counterintuitive conclusion that nuclear quantum effects harden the structure of the liquid in comparison to classical CPMD simulations at the same temperature.  Numerous studies have shown that such simulations generate an overstructured liquid~\cite{galli1,bigcpwater,martynawater}. Consequently, nuclear quantum effects would increase the discrepancy between experiment and simulation. If correct, this result would have severe implications for the accuracy of current DFT approximations of water.   

In this work we use a combination of closed and open Feynman paths to compute the pair correlation functions and the momentum distribution. We find that the liquid is significantly less structured than in computations utilizing an identical electronic structure description with classical nuclei, in qualitative agreement with experimental isotope effects and previous force field studies. The computed proton momentum distributions are also in good agreement with experiment and, unlike in empirical force field based studies, the difference between the liquid and the solid observed in experiment is reproduced. Small remaining deviations from experiment suggest some degree of over-binding in the hydrogen bond network that is likely engendered by the adopted approximate DFT description of the electronic structure. 

In the primitive discretization of the path integral formalism, the problem of describing the quantum nuclei is mapped onto a set of classical replicas coupled via harmonic interactions.  In order to compute properties such as the momentum distribution that are not diagonal in the position representation, we must use an open path to represent the nuclei for which the momentum distribution is computed~\cite{ceperley2}.  All other nuclei should be represented by closed paths.  The proton momentum distribution may then be computed from the Fourier transform of the open path end-to-end distribution.  In the present study 32 replicas are employed~\cite{burnham,morrone}.  We utilize the algorithm presented in Ref. \onlinecite{morrone} to perform the nuclear dynamics.  In this algorithm, one hydrogen per water molecule is represented by an open path, an approximation that facilitates efficient sampling with insignificant impact upon the momentum distribution.  This scheme was implemented within the CPMD~\cite{cpmd} package, which has been optimized for the IBM Blue Gene/L platform~\cite{curioni2} on which these simulations were carried out. The staging transformation~\cite{pimd1} and massive Nose-Hoover thermostat chains~\cite{nhc1} are employed in order to ensure the sustained diffusion of the trajectory, indicated by the mean square displacement of the oxygen centroid. The fictitious sampling masses are set to be 4 times larger than the corresponding physical masses of  the nuclei. 

The atomic forces are computed from first principles via the CPMD methodology~\cite{cp1}.  Electron exchange and correlation effects are treated with the BLYP functional~\cite{becke,leeyangparr}.
Troullier-Martins norm-conserving pseudopotentials~\cite{tmpp} are employed and the Kohn-Sham orbitals are expanded in a plane wave basis set with a 75 Ry cutoff. A fictitous electron mass of 340 atomic units and a time step of 3.0 atomic units is chosen~\cite{comment1}.  

The liquid water system contains 64 molecules simulated at 300K and is placed in a periodic cubic box of length $12.459 \text{\AA}$, yielding a density within 0.6\% of the experimental value at the simulation temperature. After an equilibration period of 6 ps, a production run of 12.6 ps is generated.  An electron thermostat is employed in order to maintain the fictitious electronic kinetic energy at an average value of 32K per nuclear degree of freedom.  In order to assess the effect of quantum nuclei, a CPMD simulation of liquid water with classical nuclei is also carried out utilizing otherwise identical parameters at temperatures of 300K and 330K, with average fictitous electronic kinetic energies of 17K and 18K, respectively.  A 21 ps production trajectory is generated after an 8 ps equilibration period. The proton-disordered hexagonal ice system contains 96 molecules simulated at 269K  and is modeled at experimental density in a periodic box with one side of length $13.556 \mathrm{\AA}$ and the sides in a $1: 1.15:1.09$ ratio. The electronic kinetic energy is held to a value of 27K.  After 1 ps of equilibration, a 3.8 ps production run is generated.

\begin{figure}[htp]
\vspace{-0.05in}
\begin{center}
\includegraphics[scale=.40]{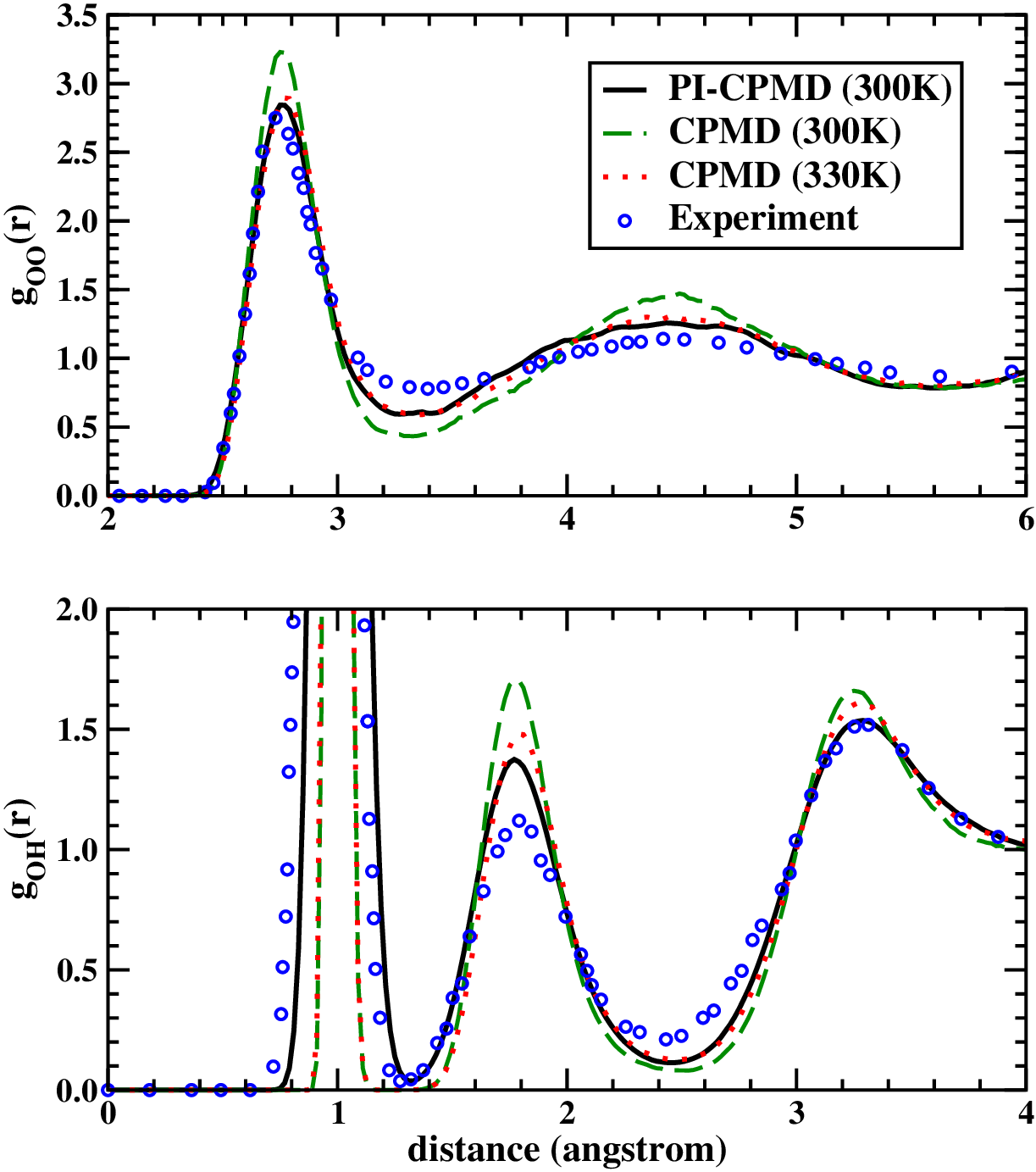}
\caption{(color online) The OO (top panel) and OH (bottom panel) RDFs in liquid water from an open PI-CPMD simulation at 300K (solid line), standard CPMD simulations at 300K (dashed line) and 330K (dotted line), and neutron scattering experiments at 298K~\cite{soper_2000} (circles) are reported.\label{fig:grs} }
\end{center}
\vspace{-0.3in}
\end{figure}

The OO and OH RDFs for liquid water are shown in Fig. \ref{fig:grs}.  They are plotted against RDFs garnered from neutron scattering experiments~\cite{soper_2000}. 
The inclusion of nuclear quantum effects leads to significantly less structured RDFs than the corresponding CPMD simulation with classical nuclei.  This qualitative observation is in agreement with previous studies of water that employ empirical potentials~\cite{martynawater}.  
The covalent peak of the OH RDF is slightly shifted, but in otherwise good agreement with the experimental neutron scattering results, though the second peak, which corresponds to hydrogen bonding interactions, is somewhat sharper, although not nearly as sharp, as the standard CPMD result.  The first peak of the OO RDF is a useful marker of the relative structuring of water.  The resultant peak from the path integral simulation is $2.84$, which is much closer to the peak height garnered from both neutron~\cite{soper_2000} and x-ray~\cite{headgordon2} scattering experiments than the peak height of $3.23$ that we obtain from a standard CPMD simulation.  The latter result is in excellent agreement with a previous study that employed a similar methodology and parameters~\cite{martynawater}. The path integral OO RDF is overall very similar to that of a standard simulation run at 330K. However, the quantum delocalization of the protons is stronger than the classical temperature effect at 330K, as indicated by the OH RDFs in Fig. \ref{fig:grs}. From these results, it appears that the overstructuring present in standard CPMD simulations at room temperature is in part mitigated by the inclusion of nuclear quantum effects.   Yet, despite the improvement, there remains some degree of overstructuring in the path integral result.  In particular, the first minimum of the OO RDF is deeper than the experimental result.

One may also evaluate the degree of structuring in the liquid from the average fraction of broken hydrogen bonds.  A hydrogen bond is defined in geometric terms by oxygen-to-oxygen and oxygen-to-hydrogen distance cutoffs that are equal to the minima of the hydrogen-bonding peaks of the RDFs (Fig. \ref{fig:grs}), and a hydrogen bond angle greater than $140$ degrees.  In the path integral representation, the distances and angles are measured from the centroid of each representative path.  Absence of broken hydrogen bonds yields a tetrahedral coordination for each molecule as in ice.  The fraction is approximately $10\%$ for liquid water~\cite{headgordon2}, and both the path integral (11\%) and standard simulation (7\%) yield averages near this value.  However, the larger number of broken hydrogen bonds in the PI-CPMD result indicate increased fluidity.

\begin{figure}[htp]
\begin{center}
\includegraphics[scale=0.28]{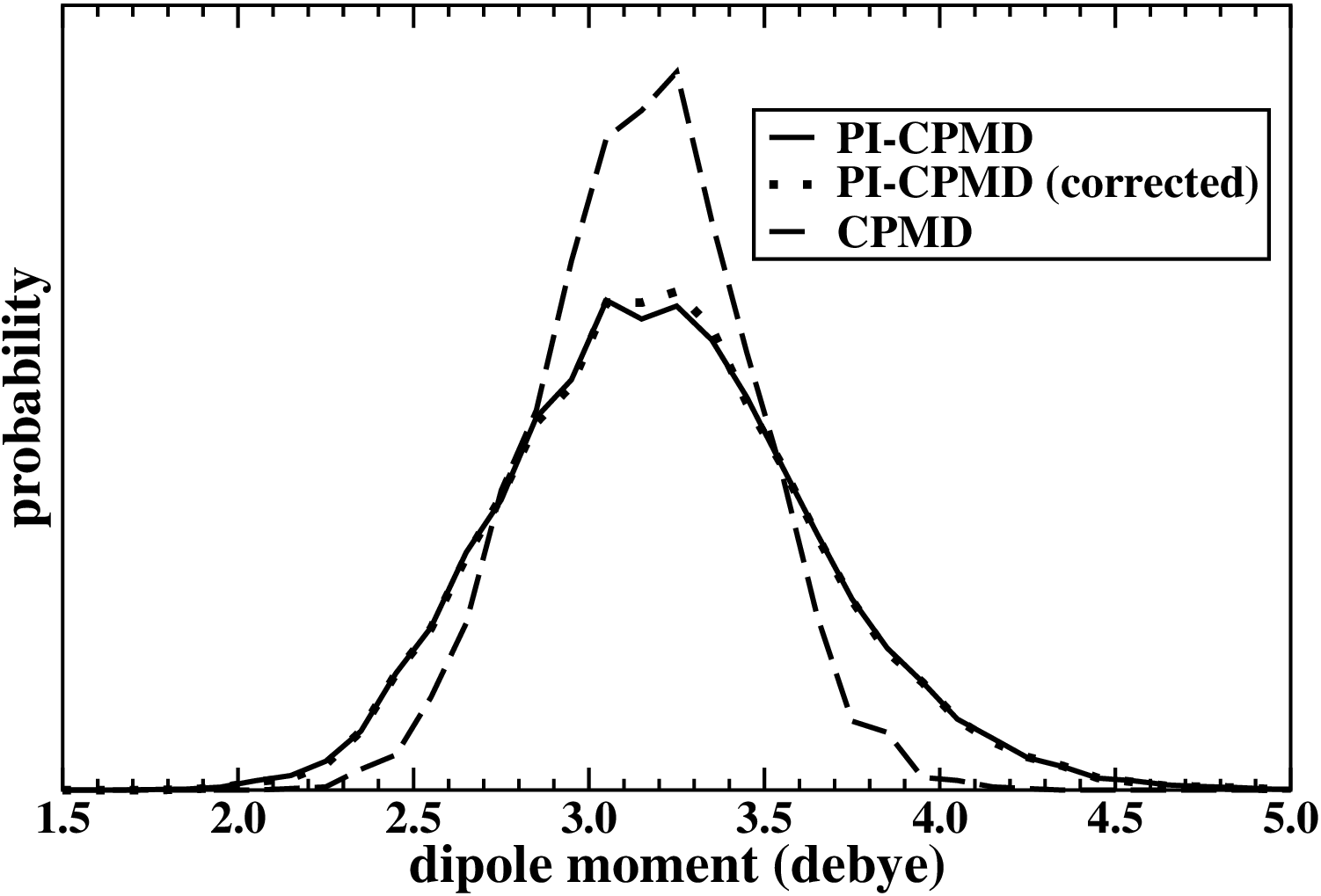}
\caption{\label{fig:dipole} The dipole moment distribution of the full PI simulation (solid line), corrected PI simulation (dotted line) and standard (dashed line) simulation of water at 300K.}
\end{center}
\vspace{-0.3in}
\end{figure}

The distribution of dipole moments in the standard and path integral simulation at 300K is depicted in Fig. \ref{fig:dipole}. The dipole moment of each individual molecule was compiled over selected configurations via the sum over ions and the centers of maximally localized Wannier functions~\cite{silvest}. 
Within the available statistics there is no appreciable change in the average dipole moment of the classical and quantal distributions. The only noticeable difference is a broadening of the path integral distribution due to the broadening of the OH covalent bond distribution evident in Fig. \ref{fig:grs}.  As a consequence, the root-mean square dipole moment is larger in the path integral than in the standard simulation. Since the dielectric constant is proportional to the fluctuation of the molecular dipole moment~\cite{manu07}, this result is consistent with the experimental finding that the dielectric constant of light water is slightly larger than that of heavy water~\cite{greenwood}.

It is important to assess how the equilibrium properties of position are affected by our ``open'' path approximation. The OH RDF is, when calculated with open path hydrogens, broader by approximately $4\%$ than when calculated with closed path hydrogens. However, if the four end replicas on each side of an open chain are excluded from the average, the corrected open OH distribution is similar to the one corresponding to closed paths, which is reported in Fig. \ref{fig:grs}. Such effects are less pronounced for the OO radial distribution function.  Open and closed hydrogen chains are used to calculate the dipole moment distribution in Fig. \ref{fig:dipole}, where a corrected plot is also reported.  One notes that the corresponding distribution is essentially the same as the uncorrected one. We also find that the open path approximation has a negligible effect on the computed fraction of broken hydrogen bonds. 

\begin{figure}[htp]
\begin{center}
\includegraphics[scale=.33]{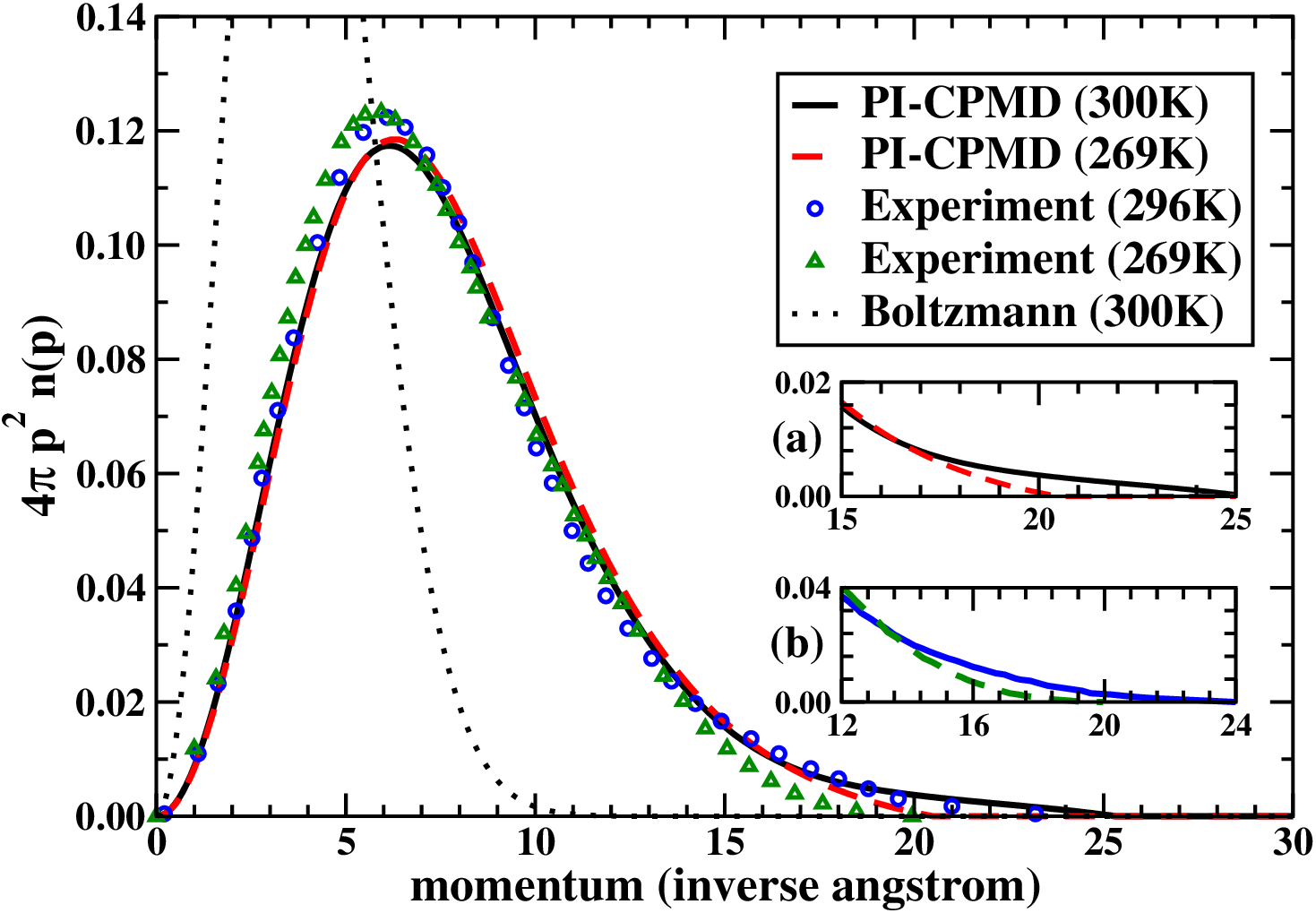}
\caption{(color online) The radial proton momentum distributions is reported in the liquid (solid line) and solid (dashed line) phases and plotted against the experimental liquid (circles) and solid (triangles) curves~\cite{reiter}, as well as the Boltzmann distribution (dashed line) at 300K. The insets (a) and (b) depict the tail of the liquid (solid line) and solid (dashed line) distributions, in the simulation and experiment, respectively.}\label{fig:pdist}
\end{center}
\vspace{-0.30in}
\end{figure}

The proton momentum distribution is computed for liquid water and ice, and is compared to neutron Compton scattering data~\cite{reiter} and the Boltzmann distribution in Fig. \ref{fig:pdist}.  There is a large distinction between the classical result, which only depends upon mass and temperature, and the actual momentum distribution which, owing to quantum effects, is sensitive to the potential energy surface.  In both phases, the proton momentum distribution of the simulation is broader than the experimental result, although the curves are generally in good agreement with each other.  Consistent with the uncertainty relation between position and momentum, the broader computed momentum distribution corresponds to the more structured (i.e. more localized) OH RDF of the simulation in comparison to experiment.  This is depicted for the liquid in Fig. \ref{fig:grs}, and is also noticed upon inspection of the ice OH RDF (not shown). The tail of the computed distribution in ice is shorter than in the liquid, which can be seen from the insets of Fig. \ref{fig:pdist}, and is in good qualitative agreement with the experimental distributions. This effect has not been reproduced in open path integral simulations that employ empirical force fields~\cite{burnham,morrone}.  Therefore the first principles potential energy surface, unlike less transferable interaction models, is sensitive to the red-shift in the OH stretch of a water molecule when it is placed in a stronger hydrogen bonding environment. 

Our position and momentum results are mutually consistent and agree with both physical intuition and the available experimental data. We find that nuclear quantum effects considerably soften the structure of the liquid and consequentially correct in large part the overstructuring present in standard first principles water. Even with the inclusion of such effects, however, there is still some degree of overstructuring present in the simulation, indicating a residual error in the description of the potential energy surface. This is likely due to the adopted treatment of exchange and correlation, as studies have shown that hybrid functionals improve the description of hydrogen bonding~\cite{hybrid_water}. It has also been suggested that the use of plane wave basis sets at typical cutoffs contributes to the overstructuring of liquid water~\cite{tucklee1}.

In this work, we have extended the PI-CPMD methodology to allow computations of the proton momentum distribution and have reported the first application of this scheme to liquid and solid water. Our results are in good agreement with neutron Compton scattering data. Given the similarity of the local environment in water and ice, the improvement provided by this approach over empirical potentials, albeit qualitatively important, is quantitatively modest.  This approach should be particularly useful in treating proton tunnelling, which involves bond breaking and forming events that are not easily captured by empirical potentials. Such events are likely to play a crucial role in proton wires that are present in biological settings~\cite{biowires}. Contributing to the understanding of such systems is a future goal of both this methodology and the related experimental techniques. 

\begin{acknowledgments}
We acknowledge support from the Fannie and John Hertz Foundation (J.M.) and NSF-MRSEC grant DMR-0213706, A. Curioni, G. Reiter, and P. Platzman for discussion, and IBM and Princeton University for the use of their computational resources.
\end{acknowledgments}

\end{document}